\documentclass[fleqn,usenatbib]{mnras}

\usepackage{newtxtext,newtxmath}

\usepackage[T1]{fontenc}

\DeclareRobustCommand{\VAN}[3]{#2}
\let\VANthebibliography\thebibliography
\def\thebibliography{\DeclareRobustCommand{\VAN}[3]{##3}\VANthebibliography}

\usepackage{graphicx}	
\usepackage{amsmath}	
\usepackage{amssymb}	

\newcommand{\teff}{$T_\mathrm{eff}$}

\title[The very massive eclipsing system WR\,21a]{
The winking eye of a very massive star: WR~21a revealed as an eclipsing binary by \textit{TESS}}
\author[R. H. Barb\'a et al.]{
Rodolfo H. Barb\'a\thanks{We dedicate this work to the memory of Rodolfo (1962--2021), who passed away during the correction process. We will miss him very much.},
Roberto C. Gamen$^{1,2}$\thanks{E-mail: rgamen@fcaglp.unlp.edu.ar},
Pablo Mart\'{\i}n-Ravelo$^{3}$,
Julia I. Arias$^{3}$,
and Nidia I. Morrell$^{4}$
\\
$^{1}$Instituto de Astrof\'{\i}sica de La Plata, CONICET--UNLP, Paseo del Bosque s/n, La Plata, Argentina.\\
$^{2}$Facultad de Ciencias Astron\'omicas y Geof\'{\i}sicas, Universidad Nacional de La Plata, Argentina.\\
$^{3}$Departamento de Astronom\'{\i}a, Universidad de La Serena, Av. Cisternas 1200 Norte, La Serena, Chile.\\
$^{4}$Las Campanas Observatory, Carnegie Observatories, Casilla 601, La Serena, Chile.
}

\date{Accepted XXX. Received YYY; in original form ZZZ}

\pubyear{2022}

\begin{document}
\label{firstpage}
\pagerange{\pageref{firstpage}--\pageref{lastpage}}
\maketitle

\begin{abstract}
WR~21a was known as a massive spectroscopic binary composed of an O2.5\,If*/WN6ha primary and an O3\,V((f*))z secondary. 
Although a minimum value, the mass estimated for the primary placed it as one of the most massive stars found in our Galaxy.
We report the discovery of photometric variations in the time series observations carried out by the Transiting Exoplanet Survey Satellite (\textit{TESS}).
These light variations are interpreted as formed by two main components: a sharp partial eclipse of the O3 secondary by the O2.5/WN6 star, and tidally excited oscillations.
Based on the light minima a new ephemeris for the system is calculated. 
The system configuration is detached and the observed eclipse corresponds to the periastron passage. During the eclipse, the light curve shape suggests the presence of the {\em heartbeat effect}.
The frequencies derived for the tidally excited oscillations are harmonics of the orbital period. 
Combining new and previously published radial velocity measurements, a new spectroscopic orbital solution is also obtained.
Using the \textsc{phoebe} code we model the \textit{TESS} light curve and determine stellar radii of $R_{\rm O2.5/WN6}=23.4$\,R$_\odot$ and $R_{\rm O3}=14.3$\,R$_\odot$ 
and an orbital inclination $i=62^\circ\!\!.2\pm0^\circ\!\!.9$. 
The latter combined with the spectroscopic minimum masses lead to absolute masses of $M_{\rm O2.5/WN6}=93.2$\,M$_\odot$ and $M_{\rm O3}=52.9$\,M$_\odot$, 
which establishes WR\,21a as belonging to the rare group of the {\em very} massive stars. 
\end{abstract}

\begin{keywords}
binaries: close --
stars: massive  -- 
stars: fundamental parameters --
stars: oscillations --
stars: Wolf–Rayet --
stars: individual: {WR 21a} 
\end{keywords}



\section{Introduction}

Although few in the Galaxy, massive stars are key astrophysical objects in view of the influence they exert on their environment.
Among them, very massive stars (VMS), i.e. those with masses above $M \gtrsim 60$\,M$_\odot$ \citep{2012ARA&A..50..107L}, form a special group. 
These rare and extreme objects defy the current scenarios of massive star formation, evolution and stellar death \citep{2015ASSL..412.....V}.
Few specimens of the VMS class exist in the Local Group, being all concentrated in massive young clusters.
Famous examples are the members of R136 at the 30~Doradus starburst \citep{2010MNRAS.408..731C},
Melnick~34 \citep{2019MNRAS.484.2692T},
R144 \citep{2013MNRAS.432L..26S,2021A&A...650A.147S},
and R145 \citep{2017A&A...598A..85S}
in the Large Magellanic Cloud,
and NGC 3603-A1 in our Galaxy \citep{2008MNRAS.389L..38S}.
VMS are representative of the top end of the initial mass function.
The masses of VMS are generally estimated using stellar atmosphere and radiative transfer codes \citep[e.g.][for R136a1, R136a2, and R136a3]{2020MNRAS.499.1918B}, or empirically determined from the simultaneous spectroscopic and photometric analysis of eclipsing binary systems \citep[e.g.][for WR\,20a]{2004ApJ...611L..33B, 2007A&A...463..981R}. 
Apart from the few existing astrometric binary systems, eclipsing binaries provide the most reliable determinations of absolute stellar masses, as well as other important parameters such as stellar radii, surface gravities, etc.

Massive stars tend to live in binary or multiple systems \citep{2012Sci...337..444S,2017IAUS..329..110S,barba2017}, most of them composing short-period binaries. 
During the last years, high-cadence space imaging surveys are revolutionising the time-domain Astrophysics and, specially, the realm of the massive stars.
The {\em Transiting Exoplanet Survey Satellite}  \citep[\textit{TESS};][]{2015JATIS...1a4003R} is a breakthrough example of such revolution. 
\textit{TESS} has revealed hundreds of new massive eclipsing and ellipsoidal binary systems \citep[e.g.][]{2020A&A...639A..81B,2021A&A...655A...4T}, opening the possibility of having a huge set of systems suitable for the calculation of precise stellar masses.
The outstanding accuracy of \textit{TESS} data has also risen the possibility of detecting brightness variations originated by dynamic tidal distortions in massive eccentric binary systems as is the case of R144 \citep{2021A&A...650A.147S}.

WR~21a (=THA 35-II-42, CDS 2134) was first identified as an H$\alpha$ emission star near the Carina Nebula region \citep{1966CoBos..35....1T,1970MmRAS..73..153W}. 
It was detected as an X-ray source (=1E 1024.0-5732) with the \textit{Einstein} satellite \citep{1989ApJ...338..338C}, subsequently identified as a Wolf-Rayet (WR) star  \citep{1994ApJ...424..943M}, and added to the {\em Catalogue of Galactic Wolf–Rayet Stars} \citep{2001NewAR..45..135V}.
\citet{2008MNRAS.389.1447N} discovered its double-lined spectroscopic (SB2) nature and presented the first orbital solution. 
Based on VLT/X-Shooter observations, \citet{2016MNRAS.455.1275T} computed minimum masses of 64.4$\pm$4.8~M$_\odot$ and 36.3$\pm$1.7~M$_\odot$ for the components of the system.
Additionally, they obtained their individual spectra through a disentangling method, and determined spectral types of O2.5\,If*/WN6ha for the primary (following the criteria defined by \cite{2011MNRAS.416.1311C}), and O3\,V((f*))z for the secondary. 
The former spectral classification is congruent with the O2 If*/WN5 presented for the star in the {\em Galactic O Star Catalog} \citep{2014ApJS..211...10S,2016ApJS..224....4M}.
Assuming for the secondary a mass according to its spectral type, they estimated an orbital inclination of $i$=58.8$\pm$2.5$^\circ$ and absolute masses of 103.6$\pm$10.2~M$_\odot$ and 58.3$\pm$3.7~M$_\odot$.
An X-ray study by \citet{2016A&A...590A.113G} including XMM-Newton, Chandra and Swift data showed that the emission  of WR\,21a in this domain exhibits small variations, except for a strengthening before periastron passage, rapidly followed by a decline as the WR star comes in front. 
These authors discarded eclipses as an explanation for this variability for several reasons, among them, the absence of eclipses in the UV range.

In this work we analyse for the first time the time-series of WR~21a obtained by the \textit{TESS} mission, and combine them with existing and new radial velocity data in order to establish precise constraints on the absolute orbital parameters and, afterwards, determine the absolute stellar masses of both components.

\section{Photometric data and light curve}

WR~21a was visited by the \textit{TESS} mission in four opportunities, two in 2019 (sectors 9 and 10), and other two in 2021 (sectors 36 and 37).
We have performed 15$\times$15 pixels cutouts (about 315$\times$315 arcsec) on the \textit{TESS} Full Frame Images (FFIs) time series \citep{2019ascl.soft05007B} centered in the position of WR~21a. 
The total exposure times were 1426~s for sectors 9 and 10, and 475~s for sectors 36 and 37.

We performed aperture photometry using the Python package \textsc{lightkurve} \citep{2018ascl.soft12013L} version 2.09 in a Python notebook. 
A stellar mask of four pixels was defined interactively in order to minimize the contamination by neighbouring sources.
The background mask was selected among the lowest brightness pixels in the cutouts, it includes about fifty pixels.
Finally, the sky background was modelled using principal component analysis, following the package recommendations.
In order to correct for low frequency variations in the time series for each sector, the extracted photometric data were normalized using a very smooth polynomial fit.

\begin{figure}
	\includegraphics[width=\columnwidth]{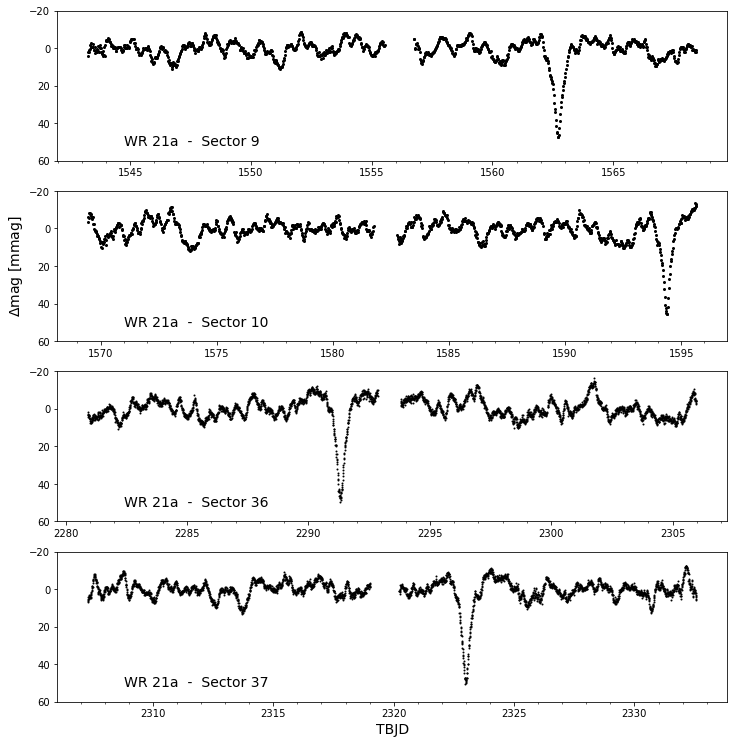}
    \caption{\textit{TESS} photometric time series obtained during sectors 9, 10, 36 and 37 (from top to bottom).}
    \label{tess}
\end{figure}

Fig.~\ref{tess} displays the resulting \textit{TESS} light curve. 
It presents a clear sharp dimming of about 5~\% in flux during each visit, which lasts for about 20.5~hours (0.027 in orbital phase). 
The overall shape resembles the light curves of the so-called {\em heartbeat} stars \citep[][]{2012ApJ...753...86T}.
In these type of binaries, during the periastron passage, proximity effects produce a noticeable change in the brightness of the system, revealed as an electrocardiogram-style pulse, which gives rise to their denomination. 
The flux minima observed in WR\,21a are periodically separated in time, in coincidence with the spectroscopic period, indicating that only one eclipse is detected in this system.
High frequency stochastic photometric variations are also present with an amplitude of about 2~\%, resembling those observed in O-type supergiants \citep[see e.g. ][]{2020A&A...639A..81B,2021A&A...655A...4T}.
Alternatively, they could be the observational signature of tidally excited oscillations \citep[TEO; ][]{1995ApJ...449..294K, 2012ApJ...753...86T}.

Since spectroscopic and photometric observations were obtained in different epochs between 2005 and 2021, we used the whole data set to improve the period determination and calculated a new ephemeris for the system.
A direct Lomb-Scargle \citep{1982ApJ...263..835S} periodogram analysis of the photometric data does not succeed at recovering the orbital period of 31.7-days. In subsection~\ref{sec:teo} we discuss in detail this periodogram.
The orbital period is approximately recovered using ``phase dispersion minimization'' \citep[\textsc{pdm};][]{1978ApJ...224..953S}, although this method does not deliver the precision needed to phase the flux minima.
Therefore, we derived the photometric ephemeris directly from the measurements of the four observed times of minima, by fitting Gaussian functions to the core of each eclipse in the \textit{TESS} time series (see Table~\ref{minima}). 
Adopting the time of minimum at sector 37 as reference epoch, the linear ephemeris for the eclipse is as follows:
\begin{equation}
\label{ephemeris}
    T_{\rm ecl}({\rm HJD})= 2\,459\,322.989\pm0.001 + 31.67855\pm0.00002\,E,
\end{equation}
being $E$, the orbital cycle.

The period derived in this photometric linear ephemeris is essentially the same as for the spectroscopic orbit, confirming that the observed dimming in the light curve is related with the orbital motion.

\begin{table}
	\centering
	\caption{Times of minima in the \textit{TESS} photometry and 
	{\em observed} minus {\em calculated} time of minima using
	the linear ephemeris of Equation \ref{ephemeris}}
	\label{minima}
	\begin{tabular}{cccc}
	\hline
	Sector & HJD & Error  & $(O-C)$\\
           & d & d    &  d\\
	\hline
09 & 2\,458\,562.7056 & 0.0015 &  0.0018\\
10 & 2\,458\,594.3801 & 0.0021 & -0.0022\\
36 & 2\,459\,291.3130 & 0.0009 &  0.0025\\
37 & 2\,459\,322.9868 & 0.0012 & -0.0022\\
\hline
\end{tabular}
\end{table}

\section{Spectroscopic data and orbital solution}

We calculated a new spectroscopic orbit using the RVs from \citet{2016MNRAS.455.1275T}, along with ten new RV measurements. 
The new epochs correspond to eight \'echelle spectra collected between 2005--2013 as part of the {\em OWN Survey} s \citep{barba2017} observing campaigns (2.5-m du Pont telescope, Las Campanas Observatory, Chile), and two spectra retrieved from the ESO database (program ID 091.D-0622(A), 2.2-m ESO/MPI telescope, La Silla Observatory, Chile).
The reader is referred to  \citet{2020MNRAS.494.3937B} for details about the reduction process. 

The RVs were derived through a double Gaussian fit of the He~\textsc{ii} $\lambda5412$ absorption line, which is clearly double in most of the spectra. 
To this aim, the dispersion and intensity of each Gaussian component was first determined from the analysis of the spectra in quadrature; then these parameters were kept as fixed, in order to fit only the line position. 
The continuum was fixed as well.
Table~\ref{dataRVs} lists the heliocentric Julian days (HJD) and RVs derived for both components, together with the instrumental configuration and resolving power $R$ of each observation.

To calculate the spectroscopic orbital solution we ran the \textsc{fotel} code \citep{2004PAICz..92....1H}, which allows the fitting of multiple RV data sets, independently determining their systemic velocities.

The calculated systemic velocities are used as zero point corrections between the different data sets. 
Considering our RV measurements we obtained $-41.5$ and $44.0$~km~s$^{-1}$ for the primary and secondary, respectively. 
The corresponding values for the RVs from \citet{2016MNRAS.455.1275T} are $-28.5$ and $32.7$~km~s$^{-1}$.
The obtained spectroscopic orbital period is $P=31.682\pm0.005$~d, which is similar to the photometric one, as we will explain below. 
Therefore, the spectroscopic orbital solution is determined using the \textsc{fotel} program, with the period fixed to the photometric value, and the remaining orbital parameters treated as free.
The resulting orbital elements and minimum masses are shown in Table~\ref{tabla_phoebe}.

\begin{table}
	\centering
	\caption{Radial velocity measurements determined from the He~\textsc{ii} $\lambda5412$ absorption line for both components in WR~21a.}
	\label{dataRVs}
\begin{tabular}{crrcc} 
\hline
HJD & RV$_\mathrm{prim}$ & RV$_\mathrm{sec}$ & instrument, telescope & $R$ \\
d & km s$^{-1}$ & km s$^{-1}$ & & \\
\hline
2\,453\,481.536	&  $-80$ &      & \'Echelle, 2.5~m, LCO  & 20\,000\\
2\,453\,489.501 & $-156$ &  229 & \'Echelle, 2.5~m, LCO  & 20\,000\\
2\,453\,490.540 & $-149$ &  249 & \'Echelle, 2.5~m, LCO  & 20\,000\\
2\,453\,491.540 & $-187$ &  256 & \'Echelle, 2.5~m, LCO  & 20\,000\\
2\,453\,772.701 & $-145$ &  194 & \'Echelle, 2.5~m, LCO  & 20\,000\\
2\,453\,875.510 &  161 & $-288$ & \'Echelle, 2.5~m, LCO  & 20\,000\\
2\,454\,200.583 &   20 &      & \'Echelle, 2.5~m, LCO  & 20\,000\\
2\,456\,435.511 & $-157$ &  245 & \'Echelle, 2.5~m, LCO  & 20\,000\\
2\,456\,468.474 & $-151$ &  231 & FEROS, 2.2~m, La Silla & 48\,000\\
2\,456\,469.483 & $-152$ &  267 & FEROS, 2.2~m, La Silla & 48\,000\\
		\hline
	\end{tabular}
\end{table}

\begin{table}
	\centering
	\caption{Orbital solution and stellar parameters.} 
	\label{tabla_phoebe}
\begin{tabular}{l  c c}
\hline\hline\noalign{\smallskip}
\multicolumn{3}{c}{Parameters obtained from the \textit{TESS} light curve}\\
\hline\noalign{\smallskip}
$P$ [d]                     & \multicolumn{2}{c}{$31.67855\pm0.00002$}\\
$T_\mathrm{ecl}$ [HJD]     & \multicolumn{2}{c}{$2\,459\,322.989\pm0.001$}\\
\hline\noalign{\smallskip}
\multicolumn{3}{c}{Orbital parameters obtained through the \textsc{fotel} code}\\
\hline\noalign{\smallskip}
Parameter & Primary & Secondary\\
\hline\noalign{\smallskip}
$T_\mathrm{periastron}$ [HJD] & \multicolumn{2}{c}{$2\,459\,323.144\pm0.001$}\\
$e$                         & \multicolumn{2}{c}{$0.695\pm$0.007}\\
$\omega$ [$^\circ$]              & \multicolumn{2}{c}{286.8$\pm$1.0}\\
$K_i$ [km s$^{-1}$]         & $158.0\pm2.7$ & $278.1\pm2.8$\\
$a_i \sin i$ [R$_\odot$]    &       71.1    & 125.2        \\
$M_i \sin^{3} i$ [M$_\odot$]&       64.6    &  36.7       \\
$q$     [M$_{2}$/M$_{1}$]   & \multicolumn{2}{c}{0.568$\pm$0.011}\\
$\mathrm{r.m.s}_\mathrm{(O-C)}$ [km s$^{-1}$]& 9.7& 10.8\\
\hline\noalign{\smallskip}
\multicolumn{3}{c}{Stellar parameters obtained by means the \textsc{phoebe} code}\\
\hline\noalign{\smallskip}
$i$ [$^\circ$]                   & \multicolumn{2}{c}{$62.19^{+0.77}_{-0.84}$}\\ 

$T_\mathrm{eff}$ [K]        &       42\,000 (fixed)             &  48\,000 (fixed)    \\
$M_i$ [M$_\odot$]           &       $93.2^{+2.2}_{-1.9}$        &   $52.9^{+1.2}_{-1.1}$    \\    
$R_i$ [R$_\odot$]           &       $23.37^{+0.52}_{-0.64}$     &   $14.28^{+0.82}_{-0.81}$    \\ 
$\log g_i$                  &       $3.669^{+0.023}_{-0.017}$   &   $3.851^{+0.048}_{-0.037}$    \\   
$\log L$ [L$_\odot$]        &       $6.18\pm0.06$               &   $6.02\pm0.09$  \\
$M_\mathrm{bol}$ [mag]      &       $-10.71\pm0.15$             &   $-10.30\pm0.22$  \\
\hline
\end{tabular}
\end{table}

\begin{figure*}
	\includegraphics[width=0.95\textwidth]{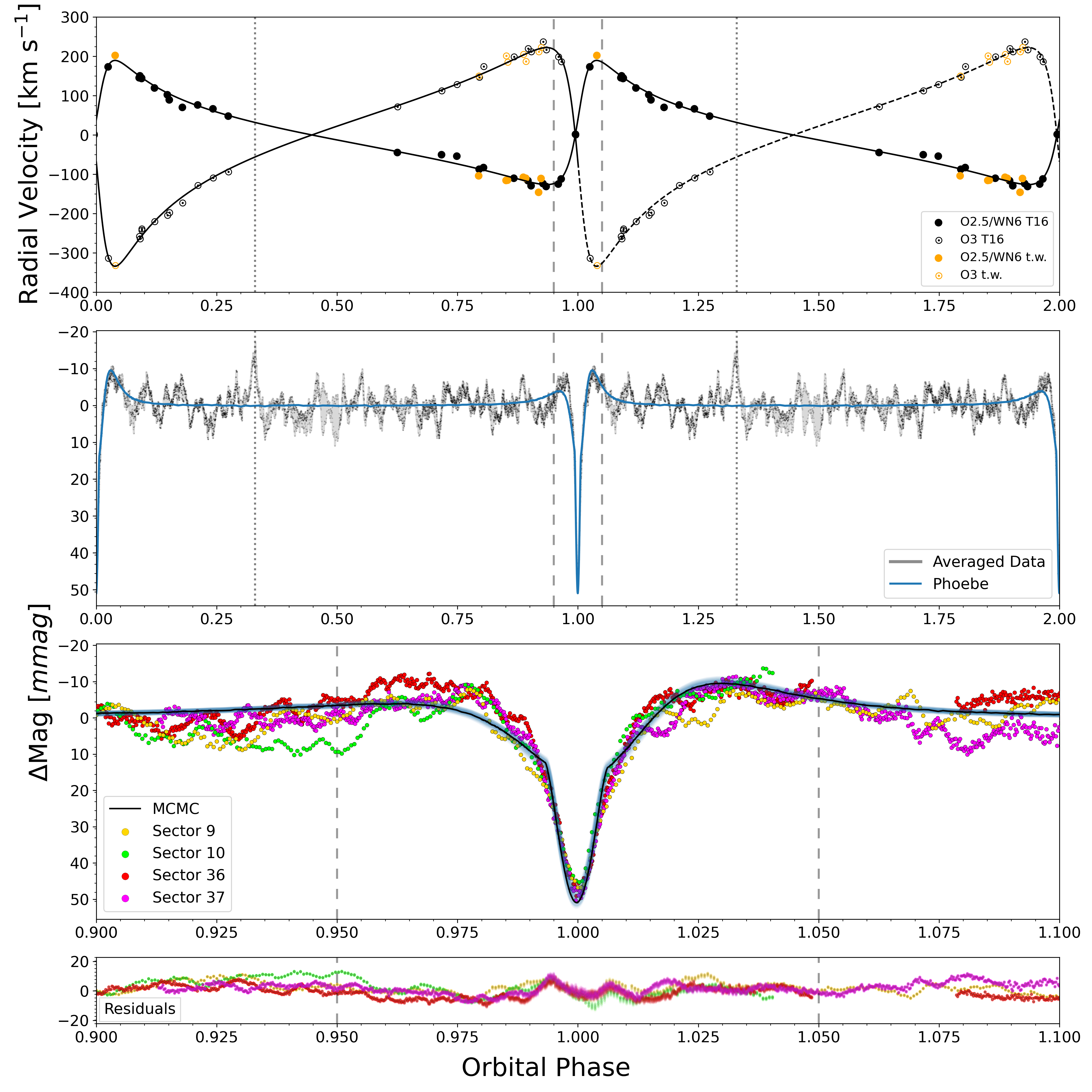}
    \caption{Top: RV curves of the WN and O components using the values of \citet{2016MNRAS.455.1275T} and Table~\ref{dataRVs} (corrected by their respective systemic velocities);
    Middle-top: Averaged \textit{TESS} light curve and the adopted \textsc{phoebe} model. 
    Middle-bottom: Comparison between the best 50 \textsc{phoebe} models (in light blue) with the photometric time series observed in the four sectors (plotted with different colours for a better visualisation). Black line represents the model for the $emcee$ solution (Table~\ref{tabla_phoebe}).
    Light curves are zoomed in the orbital phases considered in the statistical analysis of r.m.s. (i.e. $\phi$=0.95 and 1.05).
    Bottom: residuals derived from \textit{TESS} photometric observations and \textsc{phoebe} models.
    Dashed vertical lines enclose the eclipsing region that was used to calculate the r.m.s.e of the models. Dotted vertical lines represent the conjunction of the unseen eclipse (with the O3-type in the front of the system).}
    \label{fig:phoebe}
\end{figure*}

\section{WR 21a as an eclipsing binary}

We modelled the \textit{TESS} light-curve adopting the spectroscopic orbital solution previously determined, using the PHysics Of Eclipsing BinariEs (\textsc{phoebe}) package version 2.3 \citep{2005ApJ...628..426P,2016ApJS..227...29P}.
Technical details and computational procedures are described in \citet{2018maeb.book.....P}. 

We adopted the photometric ephemeris of equation~\ref{ephemeris} and performed the analysis of the system using \textsc{phoebe}. 
To calculate a binary model, some input parameters are required.
Both stellar components in WR\,21a are very hot massive stars. 
We have constrained the effective temperature (\teff) of the O2.5\,If*/WN6ha and O3\,V((f*))z components to 42\,000\,K, and 48\,000\,K, respectively.
For the O2.5\,If*/WN6ha component we adopted the \teff\ used by
\citet[][]{2004ApJ...611L..33B} for the O3\,If*/WN6 components of WR\,20a, which are spectroscopically very similar to WR\,21a.
For the O3\,V star, we adopted a \teff\ slightly larger than the O3-3.5\,V component of HD\,150136 \citep[$46\,500\pm1000$ K; ][]{2018A&A...616A..75M}, and similar to the \teff\ calculated from model atmospheres by \citet[][]{2017A&A...598A..56M} for the spectral type O3\,V (\teff$=48\,000$ K).
The \teff\ of both components in the system are outside the supported ATLAS-9 grid model atmospheres \citep{2003IAUS..210P.A20C}, hence, we used the \textsc{phoebe} option {\em blackbody} to model the atmospheres, along with the \textit{TESS} passband.
We assumed the {\em logarithmic law} to describe the limb-darkening model, using the coefficients calculated by \citet{2000A&A...363.1081C} and \citet{2017A&A...600A..30C} for the bolometric and passband limb-darkening, respectively.
We should mention that the adopted limb-darkening components correspond to models for \teff=40\,000\,K and $\log g=4.0$, because the calculated values for hotter models are suitable for sub-dwarf stars ($\log g > 4.5$), but not for massive stars.
We also assume a bolometric albedo equal to 1 for both stars. 
Orbital geometry (projected semi-axes, $a_{12} \sin i$; argument of periastron, $\omega$), and mass ratio ($q=M_\mathrm{O3}/M_\mathrm{O2.5/WN6}$) are fixed from the spectroscopic orbital solution.

As the \textit{TESS} light-curve reveals only one eclipse, it is not possible to obtain absolute values for the star radii without additional constraints on the radius of at least one component.
Absolute radius of VMS are roughly determined for a few systems, as the case of the massive twin eclipsing system WR\,20a, O3\,If*/WN6 + O3\,If*/WN6 \citep{2004ApJ...611L..33B,2004A&A...420L...9R}. 
This system is composed of stars with spectroscopic classification similar to WR\,21a, but their binary configuration is near {\em contact}, i.e. both stars are very distorted, almost filling their Roche lobes, which is not the case for WR\,21a.
Thus, we adopted as a starting point $R_1=20\,{\rm R}_\odot$, as a representative radius of the O2.5/WN6 component. 

With these assumptions, we computed different models just changing the orbital inclination from $90^\circ$ towards lower values, following a coarse step of $2^\circ$.
The first obvious conclusion is that at high inclinations ($i \sim 90^\circ$) models produce two deep eclipses. 
With $i \sim 84^\circ$, models start to show only one eclipse, being the O star eclipsed by the O2.5/WN6 component.
At these inclinations, the observed eclipse depth would only be achieved with a radius of the O-type star $R_2=4.5\,{\rm R}_\odot$, which is much smaller than what is expected \citep[$R_\mathrm{O3~V}$=13.8~R$_\odot$, c.f., ][]{2005A&A...436.1049M}.
Models improve around $i \sim 65^\circ$, when the secondary radius gets $R_2 \sim 11\,{\rm R}_\odot$.

Therefore, we have calculated $51\times21\times21 = 22491$ models in the range of $i=[50^\circ, 75^\circ]$, at steps of $\Delta i=0.5^\circ$, and stellar radii in the ranges of $R_1 = [20, 30]$\,R$_\odot$, and $R_2 = [11, 21]$\,R$_\odot$, with steps of $\Delta R_{1,2} = 0.5$\,R$_\odot$.
Fig.~\ref{fig:phoebe} (mid-top panel) shows the best model fitted to the averaged light-curve.
We calculated the root-mean-squared (r.m.s.) from the model and the observed light-curve only in the orbital phases centred in the eclipse, in the interval $\Delta\phi=[-0.05,0.05]$ (a tenth of the orbital period).
The remaining orbital phases were also modelled, but they are not used for the r.m.s. computation because the light curve is dominated by oscillations and the model is constant in flux. They are however used for the normalisation of each simulated light curve. The values of the best fit are then introduced as priors to phoebe, and the $emcee$ solver is run for the regions centred in these values ($R_1 = 23.5$\,R$_\odot$, $R_2 = 15$\,R$_\odot$, $i = 61.5^\circ$. See Fig.~\ref{arte}), with a Gaussian width of ($1.6$, $2.0$, $1.5$), respectively. The $emcee$ code was run with five walkers per parameter (for a total of 15) for 250 iterations. After the first 50 iterations were burned, with a sigma of 2 selected, $emcee$ converged to the values of radii and inclination shown in Table~\ref{tabla_phoebe}, were $\log g$ and masses were derived from these values and uncertainties were obtained by error propagation.
The eclipses observed in each of the four \textit{TESS} sectors are differently affected by such photometric variations, producing subtle changes in the depth and width.

The eclipse is zoomed in the mid-bottom panel of Fig.~\ref{fig:phoebe}, depicting the photometric time series for the four \textit{TESS} sectors. 
Superimposed are the best 50 {\sc phoebe} models. 
Their residuals are presented in the bottom panel.   

Overall, the {\sc phoebe} models describe fairly well the observations. 
However, we explored the possibility that the photometric variability during conjunctions would be produced by non--photospheric eclipses, i.e. that the light from the O3 star is scattered by free electrons in the wind of the WN component, and/or due to a pure heartbeat effect, i.e. light variations produced by the changing shape of the binary components. 

Regarding the non--photospheric eclipses, we programmed the analytical solutions provided by \citet{1996AJ....112.2227L} for the $\beta=0$ and $\beta=1$ wind laws and by \citet{2021A&A...650A.147S}, for $\beta=2$.
We explored the models for different wind parameters, such as mass--loss rate ($\dot{M}$), terminal velocity ($v_\infty$), and number of electrons per baryon ($\alpha$), and also for different stellar radius $R_\mathrm{WN}$ and orbital inclination $i$. In general, the depth of the eclipse could be reproduced (combining the appropriate parameters), but not its width.
There are two alternatives to narrow the dimming; they consist in either weakening the stellar wind (which would shrink the scattering area) or including excess emission due to wind--wind collision (WWC). 
The second possibility, following the model of \citet{2021A&A...650A.147S}, assumes that the excess emission in the WWC region is modulated by the separation between components.

\begin{figure}
	\includegraphics[width=0.99\columnwidth]{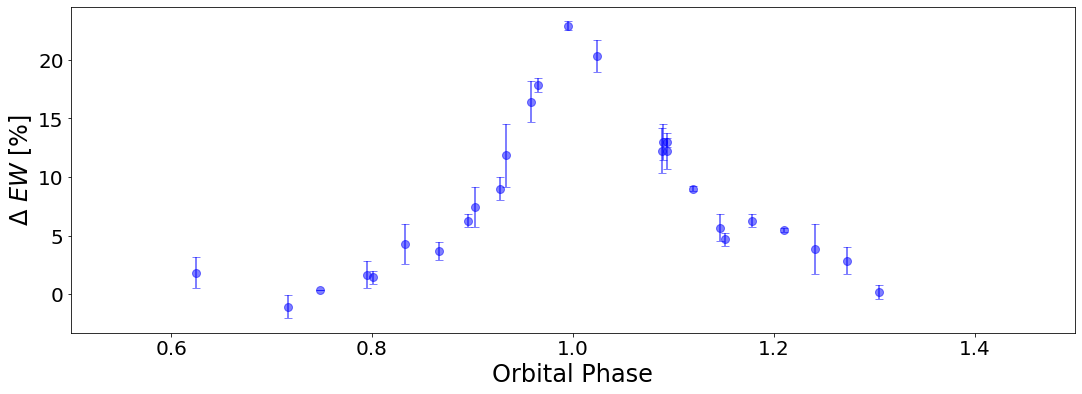} 
    \caption{Normalised equivalent width ($EW$) of the H$\alpha$ emission line measured in the 27 X-Shooter spectra (see main text).}
    \label{fig:halpha}
\end{figure}

The presence of a WWC region in this system was proved by \citet{2016A&A...590A.113G} through a tailored X-ray analysis, but its visual counterpart was not discussed. Hence, we explored how the WWC is acting on the visual wavelength range. To do this, we analysed the behaviour of the H$\alpha$ emission line.
It has been widely demonstrated that this line is a WWC indicator \citep[see e.g.][]{1997ApJ...487..380T}.
We downloaded 27 X-Shooter spectra from the Science ESO portal \citep[the ones used in the work of][]{2016MNRAS.455.1275T}, as they comprise a homogeneous dataset including H$\alpha$.
We measured the equivalent width ($EW$) of this emission line and found that it is variable (see Fig.~\ref{fig:halpha}). This variation is modulated by the orbital period, reaching its maximum during the periastron passage, being about 1.2 times stronger than at other phases.
To transform this variation in H$\alpha$ to the \textsl{TESS}-band, we performed synthetic photometry to the X-Shooter spectra. We employed the \textsc{PySynphot} tool \citep{2013ascl.soft03023S} and determined that the H$\alpha$ filter band-pass is equivalent to  1.3~\% of the \textsl{TESS}-band, and thus the 20~\% variation detected in H$\alpha$ would imply only 0.002~mag in \textsl{TESS}-band, which would be negligible or non-detectable.
Hence, including WWC to the hybrid model would not narrow the light dimming either.

\begin{figure}
	\includegraphics[width=8.5cm]{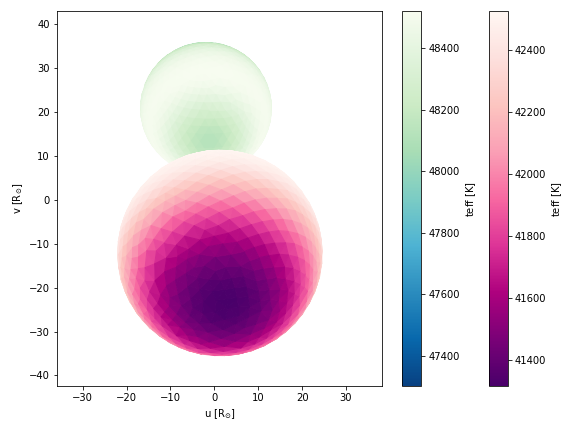} 
	\includegraphics[width=8.5cm]{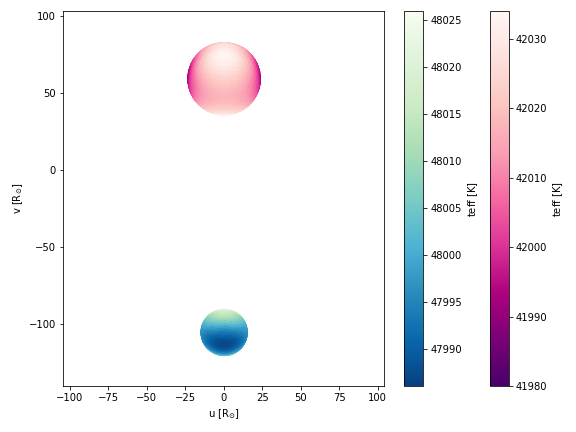}
    \caption{Geometric model of WR\,21a in the eclipse (top panel), and in the inferior conjunction (bottom panel), as seen in the plane of the sky.
    Note the change in the \teff\ of the stars during the eclipse (near periastron passage). 
    Interestingly, the tidal deformation of the O2.5/WN6 (in magenta) during that moment produces a \teff\ drop of more than 600 K in the back side of the star.
    Conversely, the O3-component (in blue) is somewhat cooler in the bulge facing the primary component but hotter in the surrounding areas.
    During the inferior conjuction, the \teff\ of the stars is very homogeneus ($\Delta$\teff$<70$ K).
    The colour maps are centred in the \teff\ adopted for each star. 
}
    \label{fig:system-config}
\end{figure}

\begin{figure}
	\includegraphics[width=8.5cm]{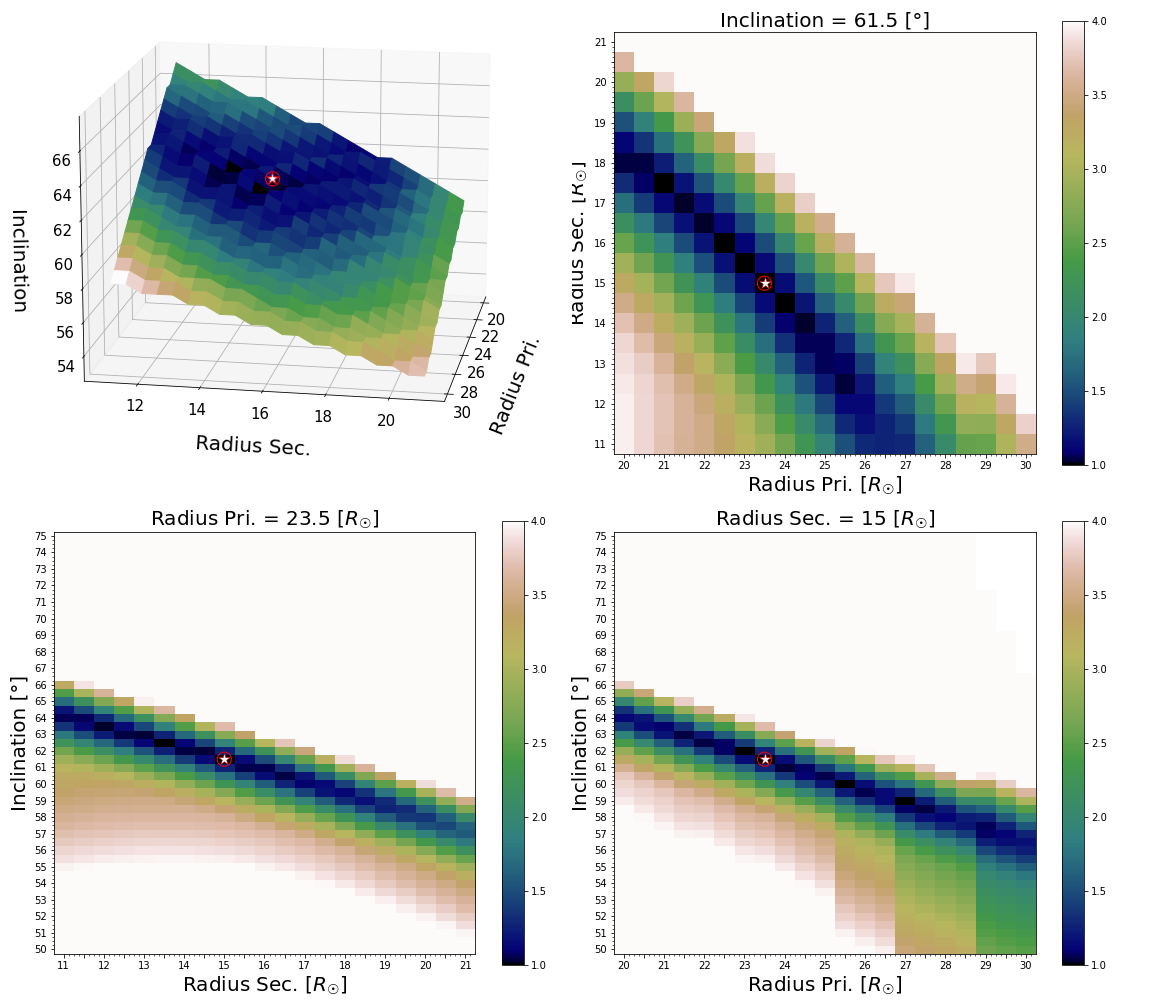} 
    \caption{3-D map distribution of normalised r.m.s. values for the set of parameters $(i,R_1,R_2)$ used to calculate the {\sc phoebe} models (upper -left). 
    The other three panels show orthogonal cuts at the position of the adopted averaged values for $R_1$, $R_2$ and $i$: 
    $(R_1,R_2)$ at fixed $i$ (upper -right), $(R_2,i)$ at fixed $R_1$ (lower-left), and $(R_1,i)$ at fixed $R_2$ (lower right).
    The colour bar represents normalised r.m.s. values. A total of 22491 models were run for the estimation of these values, which then served as priors for the $emcee$ \textsc{phoebe} solver, which provides a more precise value (Table~\ref{tabla_phoebe}), and also estimates the errors on the desired local region.}
    \label{arte}
\end{figure}

The asymmetric shape reminiscent of the {\em heartbeat} stars is recovered, although some features should be highlighted.
In the case of WR\,21a, the periastron passage is at orbital phase $\phi=0.0049$, i.e. during the eclipse of the O3 star (dotted line in Fig.~\ref{fig:phoebe}), hence any manifestation of the {\em heartbeat} effect must be observed close to those phases (i.e. eclipse). 
The shape of the eclipse core is satisfactorily modelled.
The eclipse egress is fairly modelled, too, including the characteristic {\em heartbeat} bump after the periastron passage (see Fig.~\ref{fig:system-config}), which can be explained in part as mutual irradiation effects.
The computed models show a small flux excess (2 mmag) in orbital phases before the eclipse, signature of the {\em heartbeat} effect, but a small (5 mmag) systematic deviation from observations is still present in some orbital cycles, noticeable in sector 36 during orbital phases $\phi=[0.975-0.990]$.
Another interesting feature is the bump with an amplitude of about 15~mmag located at the inferior conjuction ($\phi=0.33$). 
At this orbital phase, the stars are very well separated, $346$\,R$_\odot$ (c.f. Fig.~\ref{fig:system-config}, bottom panel), then the bump does not arise from the mutual irradiation of the stars.

The upper-left panel of Fig.~\ref{arte} shows the 3-D r.m.s. distribution of the observed and computed light-curves for the interval of selected parameters $(i,R_1,R_2)$. 
The remaining panels represent cuts in different planes, $(R_1,R_2)$, $(R_1,i)$, and $(R_2,i)$, at the position of the best solution.
It is necessary to note the model dependence, for a given inclination, with radius of the secondary component in function of the adopted radius for the primary component: smaller radii for the primary imply larger radii for the secondary, and vice versa (see r.m.s. $R_1$ vs $R_2$ panel in Fig.~\ref{arte}).

\subsection{Tidally excited oscillations}
\label{sec:teo}

Fig.~\ref{tess} shows that the \textit{TESS} light-curve of WR\,21a is ridden by a pattern of irregular variations with an amplitude up to 20~mmag.
In order to characterise these low-amplitude and high-frequency variations we removed the eclipse feature from the light-curve, and then we calculated a periodogram. 
Interestingly, the most significant frequencies  show certain coupling with the orbital period (eleven are listed in Table~\ref{lomb-scargle}, where the last column shows the ratio between pulsational and orbital frequencies). 
WR\,21a is a system composed by two very massive stars in a highly eccentric orbit, then, we could expect that strong time-dependent tidal forces may induce pulsations, a mechanism known as tidally excited oscillations (TEOs) \citep[][]{1995ApJ...449..294K}. 

As pointed out by \citet{2021A&A...647A..12K}, TEOs can be recognized in the frequency spectrum of the light curve as harmonics of the orbital frequency.
As we show in Table~\ref{lomb-scargle}, the most significant frequencies are n-times the orbital frequency, being the orbital harmonics $n=5,7,9,14,17,18,25,27,33,54$, that means all these frequencies could be related to TEOs.
\citet{2021A&A...647A..12K} proposed that the massive WN+WN+O multiple system HD\,5980 in the Small Magellanic Cloud \citep[][]{2014AJ....148...62K} is the most massive star known with a {\em heartbeat effect}, and probably with TEOs \citep[][]{2019MNRAS.486..725H} with a frequency of 3.96 d$^{-1}$. 
The present analysis shows that WR\,21a is one of the most massive systems known to present TEOs and a {\em heartbeat} system as well.
This fact is further supported by the computation of the normalised tidal potential energy, $\tilde{\varepsilon}$ \citep[see eq. 6 in ][]{2021A&A...647A..12K}, which is an indication of the 
amount of tidal deformation in the binary components.
This parameter is related to the ratio of tidal potential energy concentrated in tidal bulges at the periastron versus the gravitational binding energy of both components.
The value of $\tilde{\varepsilon}$ computed for WR\,21a is $\log \tilde{\varepsilon}= -6.7$, while it is $-7.3$ for HD\,5980.
Thus, with a total mass of 148\,M$_\odot$ ($\log M_1+M_2 = 2.17$), WR\,21a is located to the right and above of the position of HD\,5980 in \citet[][]{2021A&A...647A..12K}, figure 15.

\begin{figure*}
	\includegraphics[width=8.5cm]{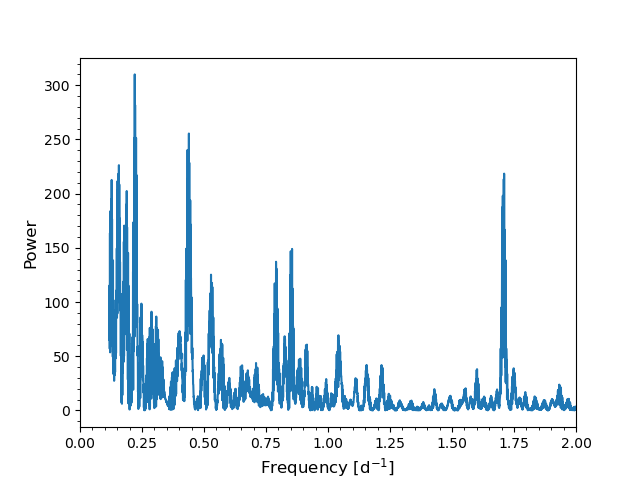}
	\includegraphics[width=8.5cm]{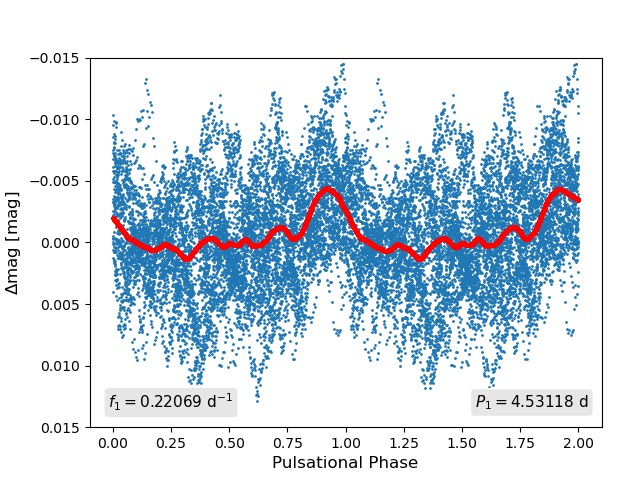}
    \caption{Left: Lomb-Scargle periodogram of {\em eclipse cleaned} \textit{TESS} light curve of WR\,21a. 
    Right: phased light-curve using the first frequency, $f_1=0.220693$ d$^{-1}$. The red line is a smoothing to the folded light curve}
    \label{fig:periodogram}
\end{figure*}

\begin{table}
	\centering
	\caption{The eleven most significant frequencies in \textit{TESS} photometric time series of WR\,21a.}
	\label{lomb-scargle}
	\begin{tabular}{ccccc}
	\hline
	\# & Frequency & Period & Power & $f_n / f_{\rm orbit}$ \\
	  & d$^{-1}$  &  d     &       &  \\
	\hline
    1  & 0.22069 &	 4.53118 &	 310 &  7.0 \\
    2  & 0.43909 &	 2.27744 &	 255 & 13.9 \\
    3  & 0.15687 &	 6.37487 &	 226 &  5.0 \\
    4  & 1.71025 &   0.58471 &   218 & 54.2 \\
    5  & 0.85512 &   1.16943 &	 149 & 27.1 \\
    6  & 0.79052 &   1.26499 &   137 & 25.0 \\
    7  & 0.52829 &   1.89289 &   125 & 16.7 \\
    8  & 0.28837 &   3.46783 &    91 &  9.1 \\
    9  & 1.04199 &   0.95971 &    69 & 33.0 \\
    10 & 0.27991 &   3.57263 &    68 &  8.9 \\
    11 & 0.56828 &   1.75969 &    65 & 18.0 \\
\hline
\end{tabular}
\end{table}

\section{Concluding remarks}

We unveiled four eclipses in the very massive binary system WR~21a using photometric time series collected by the \textit{TESS} mission.
The overall shape of the light curve during the eclipse phases mimics that of {\em heartbeat} systems.
We fitted binary models to the light-curve by means of the \textsc{phoebe} code.
Models were built based on a grid of radii for both massive components, and inclinations, keeping fixed the spectroscopic orbital elements and mass ratio.
We have determined the best model parameters, $R_1$, $R_2$ and $i$, by means the $emcee$ solver implemented to the \textsc{phoebe} code.
The orbital inclination, $i=62^\circ\!\!.2\pm0^\circ\!\!.9$ is the key parameter to calculate the absolute stellar masses of the system.
This determination of the system inclination is in good agreement with the one obtained by \citet{2016MNRAS.455.1275T}, assigning the mass calibrated by \citet{2005A&A...436.1049M} of
58.3~M$_\odot$ to the O3~V secondary.

We ruled out the possibility that the discovered eclipses are produced by scattering of light from the O-type component by free electrons in the wind of the WN star. At least with the presently available models, it was not possible to reproduce the eclipse widths under such assumption. 

The massive O3\,V((f*))z companion is characterised as a star with an absolute mass of $M_2 = 53$~M$_\odot$, and a radius of 14.3~R$_\odot$. 
There are only two Galactic O3-type stars for which absolute mass determinations are available; those are the O3.5\,V components of the triple system HD~150136 and the O3.5\,V primary component of the SB2 system HD~93205.
For HD\,150136, \citet{2013A&A...553A.131S}, using the PIONIER combiner at the Very Large Telescope Interferometer (VLTI), derived the three-dimensional orbit of the outer system, which combined with the RV curves of the inner pair, allowed to calculate a mass of 62.6$\pm$10~M$_\odot$ for the O3.5 component. 
In the case of HD\,93205, a high-quality spectroscopic solution was determined by \citet[][]{2001MNRAS.326...85M}, which was used to derive absolute masses based on apsidal motion analysis \citep[][]{2002MNRAS.330..435B}. The mass determined for the O3.5 component is 60$\pm$19~M$_\odot$.
Therefore, our empirical mass determination for the O3-component in WR\,21a is comfortably located in the mass range of similar stars in massive systems.

We have obtained a mass of 93.2$\pm$2.2~M$_\odot$ for the  O2.5\,If*/WN6ha component, confirming it as a VMS.
We must remark that reliable dynamical masses of Galactic VMS, i.e. determined through the spectroscopic and photometric calculation of orbital parameters in eclipsing systems, are very scarce. 
Among the Galactic systems, we bring up NGC 3603-A1 \citep[116+89~M$_\odot$;][]{2008MNRAS.389L..38S}, WR~20a \citep[82+83~M$_\odot$;][]{2004ApJ...611L..33B}, and Arches F2 \citep[WR 102aa, 82+60~M$_\odot$;][]{2018A&A...617A..66L}.
These three massive systems present light curves compatible with near-contact eclipsing SB2 binaries, in contrast to the case of WR~21a, which is clearly a detached system.
The radius of the O2.5\,If*/WN6ha component is slightly larger than that measured for the components of the eclipsing system WR\,20a. 
If WR\,21a was ejected from the massive star cluster Westerlund\,2 as proposed \citep[][]{2011MNRAS.416..501R}, both WR\,21a and WR\,20a should have the same age, and thus the radius difference would be related to the mass difference, being the former more massive.

This discovery places WR~21a as a new benchmark for the evolutionary analysis of VMS. 
Its light curve indicates that both components are well detached, which is also confirmed by the computed star radii. 
Given the youth of these stars, they are still evolving independently, before the first mass-transfer stage.
The presence of tidally excited oscillations opens the possibility of studying in detail the relationship of tidal oscillations and internal structure of the stars. 

\section*{Acknowledgements}

We thank the anonymous referee whose comments and suggestions have helped to improve this work.
RCG acknowledges support from grant PICT 2019-0344.
Also, we thank the
director and staff at LCO  for the use of their facilities and  kind hospitality during the observing runs.
RHB, RCG and NIM thank Diego Armando Maradona (Barrilete Cósmico), for brightening up their lives.

Software. This research made use of Lightkurve, a Python package for Kepler and \textit{TESS} data analysis (Lightkurve Collaboration, 2018) (\url{https://docs.lightkurve.org/index.html}), and ''PHysics Of Eclipsing BinariEs'' (\textsc{phoebe}) package ( \url{http://phoebe-project.org}).

\section*{Data Availability}

This paper is based on public data of the \textit{TESS} mission, and spectroscopic data belonging to the {\it OWN Survey} team, available on reasonable request to Dr. Roberto Gamen, the public ESO database, and published radial velocities \citep[][]{2016MNRAS.455.1275T}.


\bibliographystyle{mnras}
\bibliography{own}

\begin{appendix} 
\section{Simple heartbeat model}

We tried to reproduce the light curve of WR~21a programming the analytic model for the tidal  oscillations induced in eccentric binaries \citep{1995ApJ...449..294K}, which was successfully applied by \citet{2012ApJ...753...86T} on several systems. 

As it is shown in \citet{2012ApJ...753...86T} the light variations just depend on some of the orbital parameters, namely $e$, $\omega$ and $i$. In our case the two former are determined from the RV orbital solution. Then, we only considered variations in the orbital inclination and the amplitude scaling factor.

\begin{figure}
	\includegraphics[width=0.99\columnwidth]{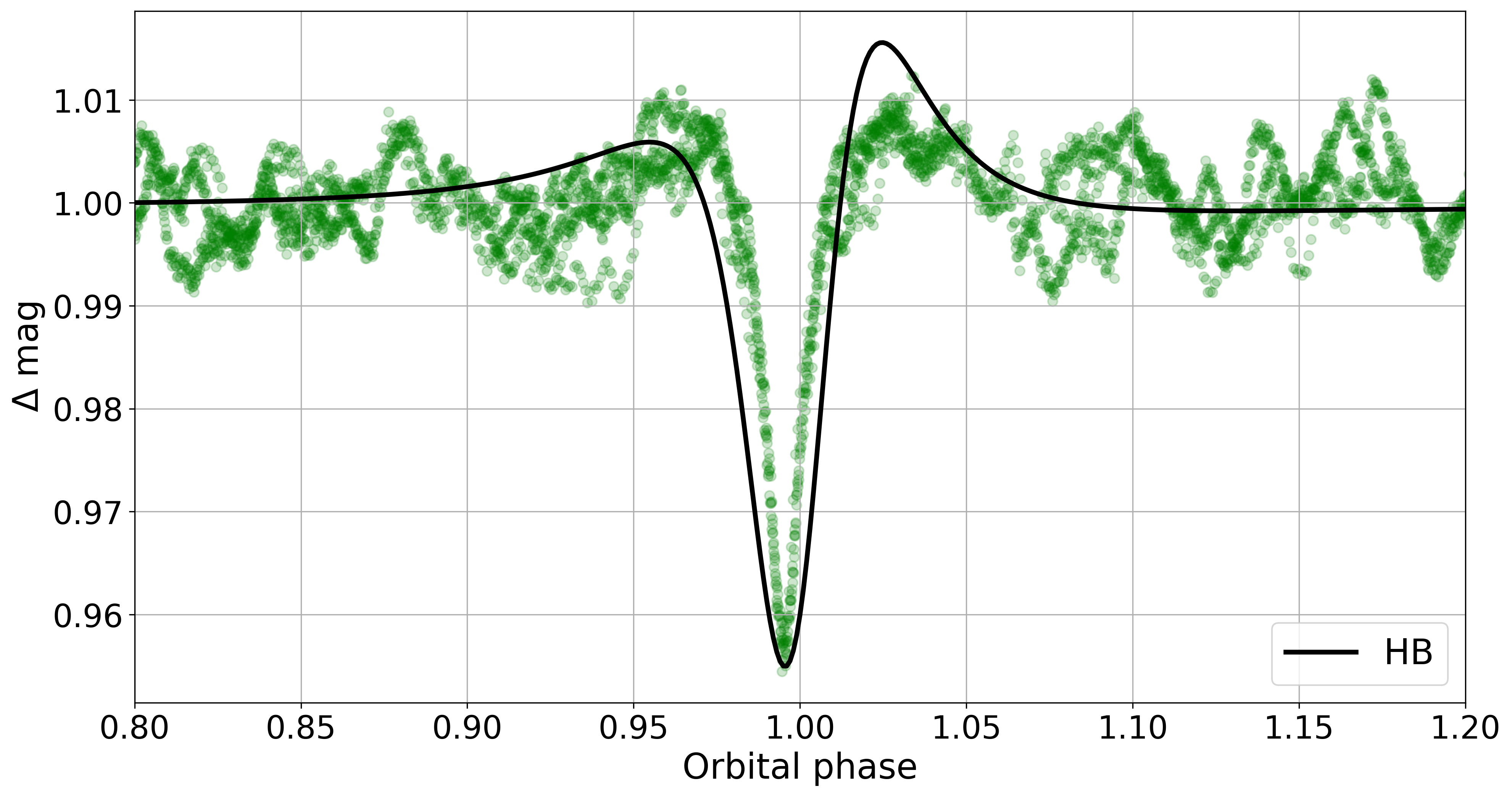} 
    \caption{Light curve of WR~21a and the heartbeat model.}
    \label{fig:HB}
\end{figure}

The major problem is clearly related to the fitting of the eclipse width. A reasonable fit was obtained with the same inclination of the \textsc{phoebe} model, but as can be noted in the Fig.~\ref{fig:HB}, the light dimming predicted is broader than observed. Any other combinations also resulted in broader dimmings.
As this simple model just considers the change in shape of the star, perhaps most sophisticated models, such as \textsc{gyre-tides} \citep{2021AAS...23743804S}, could better fit the observed light curve, but this is out of the scope of this work.

\section{Hybrid model}

We also tried to reproduce the observed light curve of WR~21a programming the analytical solutions for non-photospheric eclipses developed by \citet{1996AJ....112.2227L} for the $\beta=0$ and $\beta=1$ wind laws and by \citet{2021A&A...650A.147S}, for $\beta=2$, plus the excess emission due to wind--wind collisions (WWC) proposed by \citet{2021A&A...650A.147S}.

\begin{figure}
	\includegraphics[width=0.99\columnwidth]{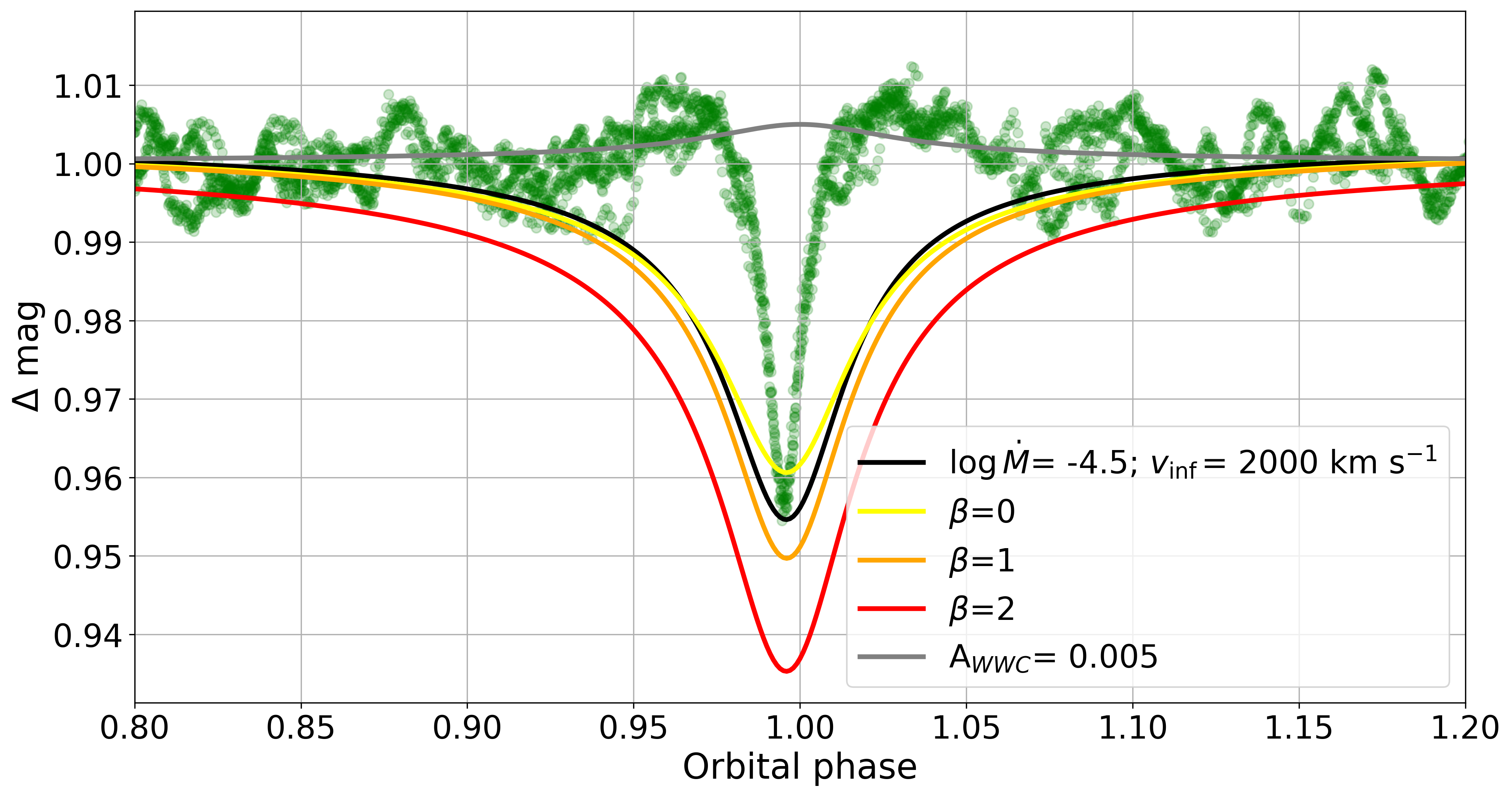}
	\includegraphics[width=0.99\columnwidth]{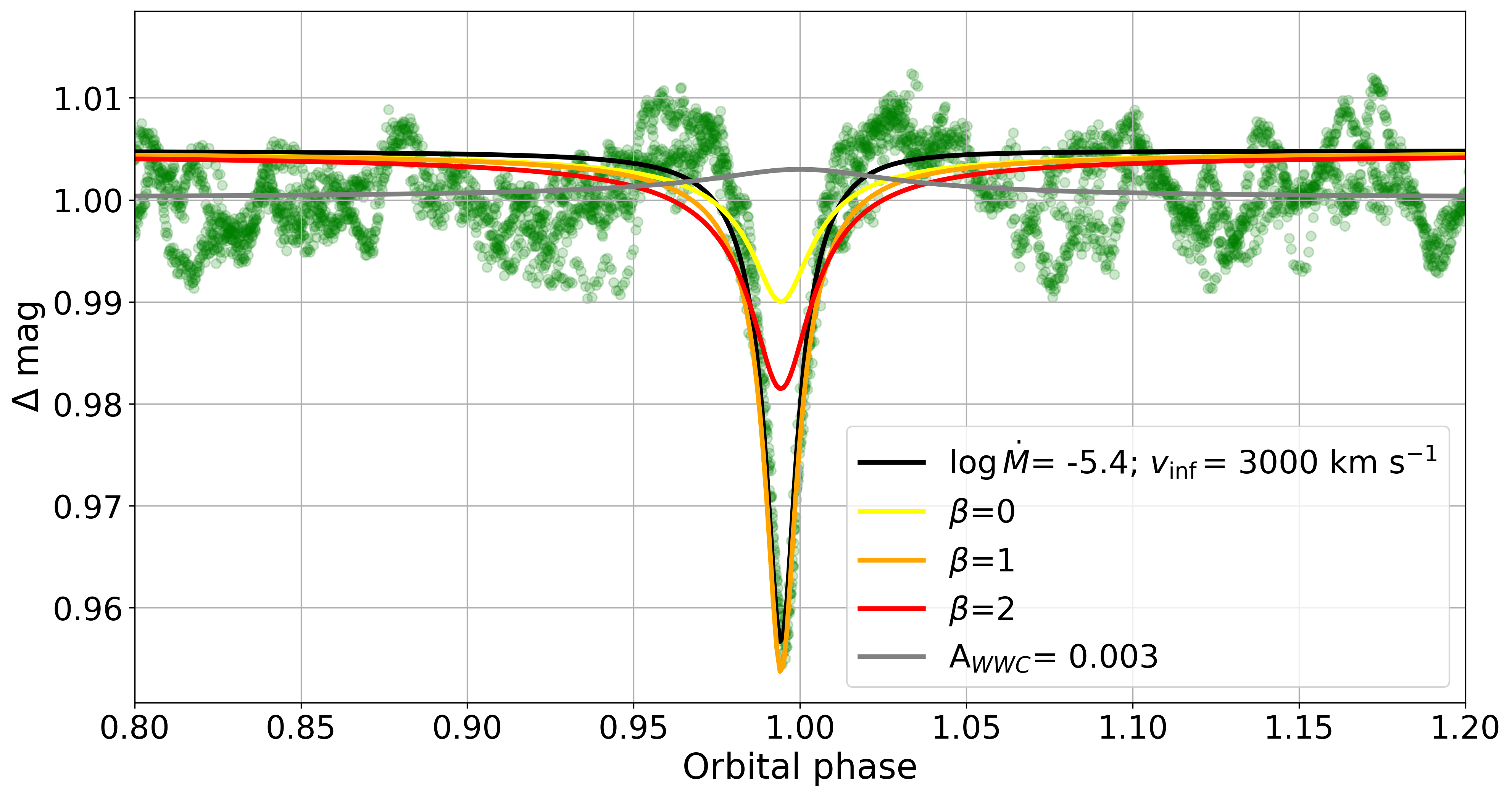}
    \caption{Light curve of WR~21a and some hybrid models.}
    \label{fig:L96}
\end{figure}

We explored the non-photospheric eclipse models altering some of the wind parameters, such as the mass--loss rate ($\dot{M}$), the terminal velocity ($v_\infty$), and the number of electrons per baryon ($\alpha$), as also the stellar radius $R_\mathrm{WN}$, and the orbital inclination $i$. In general, the depth of the eclipse could be reproduced (combining the appropriate parameters), but not its width (see the top panel of Fig.~\ref{fig:L96} for some examples).

There are two alternatives to narrow the light dimming. These are either weakening the stellar wind (smaller $\dot{M}$ and increasing its $v_\infty$), or adding a WWC region.
The latter assumes that the excess emission in the WWC is modulated by a certain power of the separation between stars described by the parameter $\gamma_\mathrm{WWC}$.
This contribution, when combined to the non--photospheric model, tends to narrow the light dimming. However, as it is not centred at the phases where the eclipses do occur but at the periastron passage, it introduces an asymmetry in the light curve, which is not seen in the observations. 

The best model found with this procedure implies a very weak wind, characterised by a mass-loss rate $\dot{M}=-5.4$ and a terminal velocity $v_\infty=3000$~km~s$^{-1}$ which seem unrealistic.

We conclude that this hybrid model is not able to reproduce the narrow eclipses observed in the WR~21a light curve.

\end{appendix}

\bsp	
\label{lastpage}
\end{document}